\documentclass[%
 reprint,
superscriptaddress, groupedaddress,
showpacs,preprintnumbers, nofootinbib, nobibnotes,
 amsmath,amssymb,
prb,
]{revtex4-1}

\usepackage{graphicx}
\usepackage{dcolumn}
\usepackage{bm}
\usepackage{amssymb}   
\usepackage{amsmath}
\usepackage{color}
\usepackage[normalem]{ulem}
\usepackage{siunitx}
\usepackage{mathrsfs}
\usepackage{tikz}
\usepackage{amsfonts}
\usepackage{kantlipsum}
\usepackage{subfigure}
\usepackage{multirow}
\usepackage{esint}

\begin{document}


\title{Plasmon-induced nonlinearity enhancement and homogenization of graphene metasurfaces}

\author{Jian Wei You}
\author{Nicolae C. Panoiu}
\affiliation{ Department of Electronic and Electrical Engineering, University College London, London WC1E 7JE, United Kingdom.  Email: n.panoiu@ucl.ac.uk
}%

\date{\today}

\begin{abstract}
We demonstrate that the effective third-order nonlinear susceptibility of a graphene sheet can be
enhanced by more than two orders of magnitude by patterning it into a graphene metasurface. In
addition, in order to gain deeper physical insights into this phenomenon, we introduce a novel
homogenization method, which is subsequently used to characterize quantitatively this nonlinearity
enhancement effect by calculating the effective linear and nonlinear susceptibility of graphene
metasurfaces. The accuracy of the proposed homogenization method is demonstrated by comparing its
predictions with those obtained from the Kramers-Kronig relations. This work may open up new
opportunities to explore novel physics pertaining to nonlinear optical interactions in graphene
metasurfaces.
\end{abstract}

\maketitle


Since it has been first isolated from graphite \cite{ngmj04sci}, the unique and striking properties
of graphene have spurred intense research efforts to develop and synthesize new two-dimensional
(2D) materials. So far, graphene and other 2D materials have already had a great impact both as
facilitators of key advancements in fundamental research, as well as enablers of new devices
operating in a broad spectrum, ranging from ultraviolet, visible and down to microwave frequencies
\cite{xhdm14natpho,ylzw13natphot,fbip14natnano,lcck17natmat,g09sci,krz14natcom,zxy16nanoenergy}.
For instance, due to their unique linear physical properties, 2D materials have found important
applications to electronics \cite{g09sci,fbip14natnano}, sensors \cite{krz14natcom}, and solar
cells \cite{zxy16nanoenergy}. Equally important, the nonlinear optical properties of graphene have
facilitated the development of new active photonic devices with improved functionality
\cite{jkws16nanolett,lccg11natphot,nbng13lpr,dymc15prb} and the exploration in new physical
conditions of fundamental phenomena, including spatial solitons \cite{nbng13lpr} and tunable Dirac
points \cite{dymc15prb}.

Inspired by the concept of metasurfaces, the research in graphene has expanded from the study of
uniform monolayer configurations to nanopatterned surface structures. These graphene metasurfaces
have open up promising new routes towards photonic devices with specially engineered linear and
nonlinear optical responses \cite{ybbp19Nano,al13oe,yywp17prsa,zfzz15oe,yp18oe}. A key role in
these developments has been played by versatile and powerful numerical and analytical methods. In
particular, the use of homogenization methods that reduce the optical response of a metasurface to
that of a homogeneous layer of material characterized by specific optical constants is ubiquitous
in the design process and analysis of metasurfaces. Two of the most common homogenization methods
are the scattering-parameter approach \cite{ssmc02prb} and the field-averaging procedure
\cite{sp06josab}. To date, they have been applied mostly to metasurfaces containing linear,
dispersive, and isotropic materials, with scarce efforts being devoted to the nonlinear case
\cite{ls10oc,gvlk16prb}. One of the main reasons for this is that the high-order nonlinear
susceptibilities of graphene are anisotropic, which significantly hinders the extension of the
existing linear homogenization methods to the nonlinear case.

In this Letter, we propose a novel linear and nonlinear homogenization method for metasurfaces
containing graphene. The challenges arisen from the nonlinear and anisotropic characteristics of
such metasurfaces are overcome by introducing a set of auxiliary physical quantities, which allows
one to unambiguously match the far-field optical response of the metasurface with that of a
homogeneous layer of material with certain constitutive parameters. This novel homogenization
method is used to study the linear and nonlinear optical properties of a generic graphene
metasurface. Our analysis reveals that, at the resonance frequencies of surface plasmons of
nano-sized graphene constituents of the metasurface the effective third-order susceptibility of the
metasurface is enhanced by more than two orders of magnitude as compared to that of a graphene
sheet.

\begin{figure}[t!]
\centering
\includegraphics[width=8.5 cm]{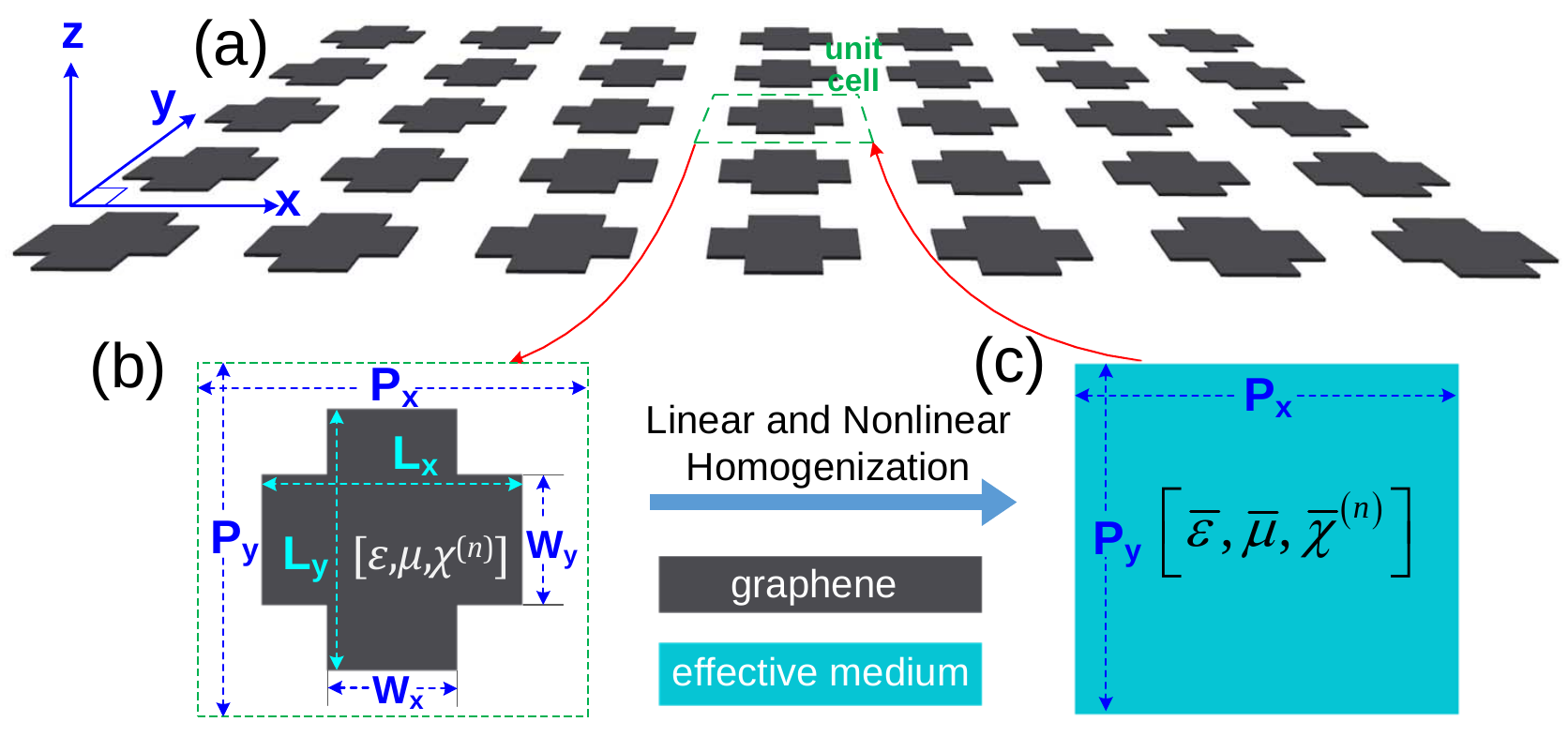}
\caption{(a) Schematics of a graphene-based metasurfaces consisting of a rectangular array of
graphene cruciform patches. (b) Geometry and material parameters size of a unit cell of the
metasurface. (c) The homogenized metasurface, characterized by effective linear and nonlinear
optical constants.} \label{fig:HomoSchematic}
\end{figure}

The schematics of the graphene metasurface investigated in this work is presented in
Fig.~\ref{fig:HomoSchematic}(a), with the unit cell depicted in Fig.~\ref{fig:HomoSchematic}(b).
The metasurface lies in the $x-y$ plane, and consists of a rectangular array of cruciform graphene
patches. The symmetry axes of the array and cruciform patches are along the $x$- and $y$-axes, the
corresponding periods being $P_x$ and $P_y$. The length and width of the arms of the crosses are
($L_x$, $L_y$) and ($W_x$, $W_y$), respectively. Unless otherwise specified, the values of these
parameters are $P_x=P_y=\SI{200}{\nano\meter}$, $L_x=L_y=\SI{180}{\nano\meter}$, and
$W_x=W_y=\SI{75}{\nano\meter}$. Moreover, the relative electric permittivity of graphene is
$\varepsilon_{g}=1+i\sigma_s/(\varepsilon_0\omega h_{g})$, where $h_{g}=\SI{0.5}{\nano\meter}$ is
the thickness of graphene, $\omega$ is the frequency, and the graphene surface conductivity,
$\sigma_s$, is described by the Kubo's formula \cite{gp16book}:
\begin{equation}\label{eq:sigma}
\sigma_{s} = \frac{e^2 k_B T\tau}{\pi\hbar^2\overline{\omega}}\left[\frac{\mu_c}{k_B T} +2\ln
\left(e^{-\frac{\mu_c}{k_B T}} + 1\right) \right]
+\frac{ie^2}{4\pi\hbar}\ln\frac{\xi-i\overline{\omega}}{\xi+i\overline{\omega}}.
\end{equation}
Here, $\mu_c$, $T$, and $\tau$ are the chemical potential, temperature, and relaxation time,
respectively, $\overline{\omega}=1-i\omega\tau$, and $\xi=2\vert\mu_c\vert\tau/\hbar$. In this
study, we use $\mu_c=\SI{0.2}{\electronvolt}$, $\tau=\SI{0.1}{\pico\second}$, and
$T=\SI{300}{\kelvin}$.

Generally, the linear constitutive parameters of a homogenized metasurface are uniquely determined.
In the nonlinear case, however, the effective nonlinear susceptibilities of a homogenized
metasurface are not uniquely determined as in general there are more independent components of
these tensor quantities than available constitutive relations. Our homogenization method
circumvents this problem by introducing a set of auxiliary variables that define the linear and
nonlinear polarizations. These auxiliary variables, and subsequently the linear and nonlinear
effective optical constants of the homogenized metasurface, are determined by requiring that the
averaged linear and nonlinear polarizations in the original and homogenized metasurfaces are
\textit{termwise} identical.

The homogenization method presented here consists of two steps: First, a standard field averaging
method \cite{sp06josab} is used to determine the averaged fields at the fundamental frequency (FF)
and the effective electric permittivity. In the second step, the fields at the FF are used to
determine the nonlinear polarization and, through a set of auxiliary variables, the effective
nonlinear susceptibility of the homogenized metasurface. As nonlinear interaction we choose the
third-harmonic generation (THG), but our method can be readily used to study other nonlinear
interactions characterized by nonlinear polarizations than can be expressed in terms of the field
at the FF.

To begin with, let us consider the constitutive relation in a linear and anisotropic material that
relates the flux density and the electric field, $D_i=\sum\nolimits_j \varepsilon_{ij}E_j$, where
$\varepsilon_{ij}$ is the permittivity, and $i,j=x,y,z$. The spatial average of the fields
$\mathbf{E}$ and $\mathbf{D}$ is defined as:
\begin{subequations}\label{eq:AverF}
\begin{align}
\label{eq:AverE}\overline{\mathbf{E}}(\omega)&=\frac{1}{V}\int_V\mathbf{E}(\mathbf{r},\omega)d\mathbf{r}, \\
\label{eq:AverD}\overline{\mathbf{D}}(\omega)&=\frac{1}{V}\int_V\mathbf{D}(\mathbf{r},\omega)d\mathbf{r},
\end{align}
\end{subequations}
where $V$ is the volume of the unit cell. These averaged fields can be used to define the effective
permittivity of the homogenized metasurface, $\overline{\varepsilon}_i
=\overline{D}_i/\overline{E}_i$.

This approach is only applicable to metasurfaces containing isotropic materials, whose permittivity
tensor is diagonal, but it can be readily extended to the more general case of anisotropic
structures by introducing a new auxiliary quantity, $d_{ij}=\varepsilon_{ij}{E_j}$. Then, the
constitutive relation for anisotropic materials is expressed as $D_i = \sum\nolimits_j {d_{ij}}$.
Imposing now the condition that the averaged flux density in the metasurface and the flux density
in the homogenized metasurface are \textit{termwise} identical, the effective permittivity tensor
can be calculated as:
\begin{equation}\label{eq:AverEpsilon}
\overline{\varepsilon}_{ij}(\omega ) = \overline{d}_{ij}(\omega)/\overline{E}_j(\omega),
\end{equation}
where the averaged auxiliary quantity is given by:
\begin{equation}\label{eq:Averdij}
\overline{d}_{ij}(\omega) =
\frac{1}{V}\int_V\varepsilon_{ij}(\mathbf{r},\omega)E_j(\mathbf{r},\omega)d\mathbf{r}.
\end{equation}

The thickness of graphene, and more generally of 2D materials, is much smaller than the optical
wavelength at infrared and THz frequencies, and thus one can assume that the optical field is
uniform across graphene. Therefore, the volume integrals in \eqref{eq:AverF} and \eqref{eq:Averdij}
become surface integrals over the mid-section plane of graphene patches.
\begin{figure}[t!]
\centering
\includegraphics[width=8.5 cm]{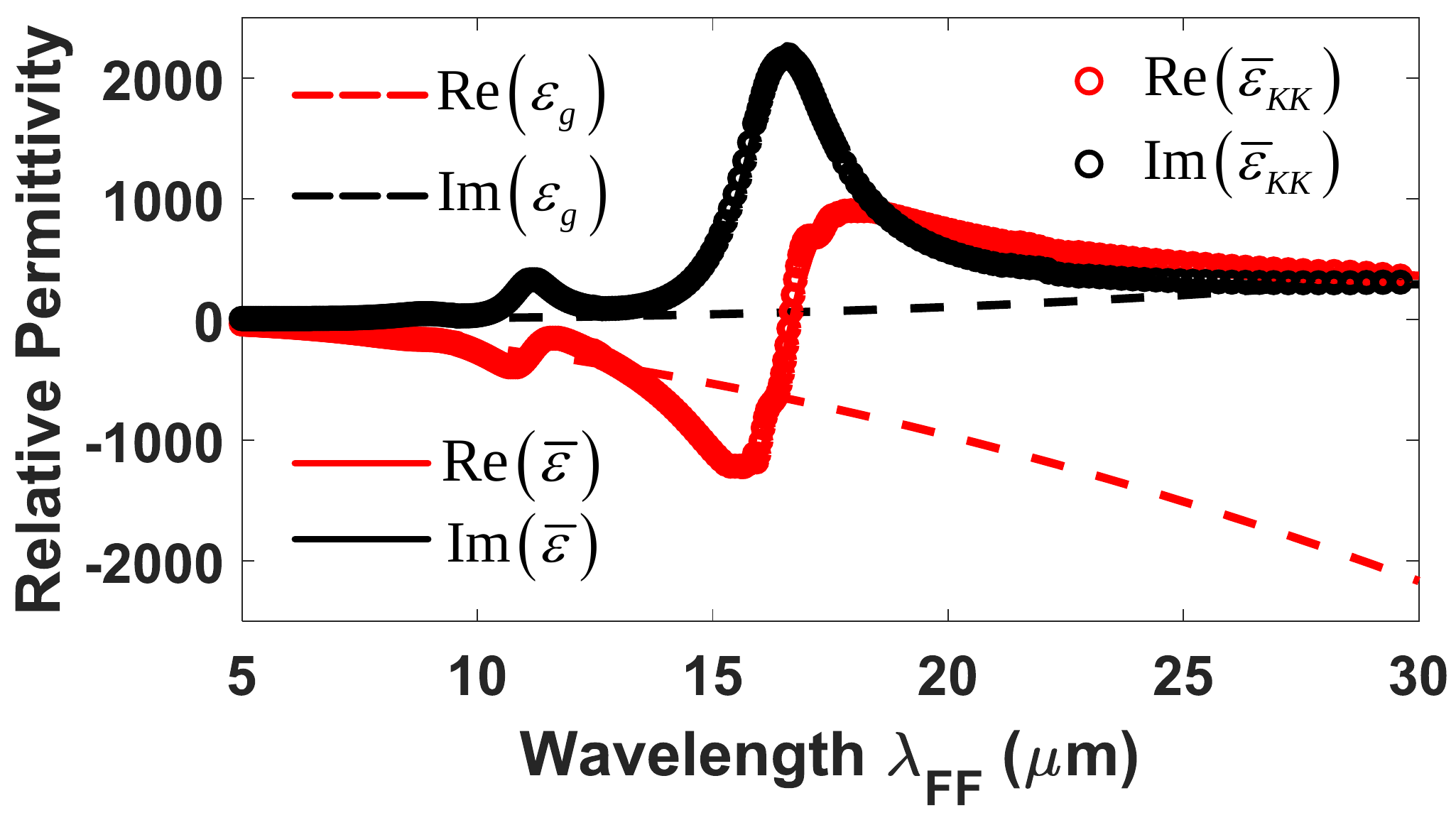}
\caption{Wavelength dependence of the real and imaginary parts of graphene permittivity,
$\varepsilon_{g}$, the effective permittivity of the homogenized cruciform metasurface,
$\overline{\varepsilon}$, and the corresponding permittivity, $\overline{\varepsilon}_{KK}$,
calculated from $\overline{\varepsilon}$ using the Kramers-Kronig relations.}
\label{fig:HomoSuscepbilityFF}
\end{figure}

In order to validate our proposed linear homogenization approach, a graphene cruciform metasurface
(see Fig.~\ref{fig:HomoSchematic}) has been studied using the FDTD method. In the numerical
simulations, the graphene cruciform metasurface is illuminated by a normally incident plane wave
linearly polarized along the $x$-axis. The relevant results are presented in
Fig.~\ref{fig:HomoSuscepbilityFF}, where we plot the wavelength dependence of the intrinsic
relative permittivity of graphene, the effective permittivity of the homogenized graphene
metasurface retrieved from \eqref{eq:AverEpsilon}, and the corresponding permittivity determined
from the Kramers-Kronig (KK) relations \cite{lspv05book}, $\overline{\varepsilon}_{KK}$:
\begin{subequations}\label{eq:KK}
\begin{align}\label{eq:ReKKR}
\mathfrak{Re}\left\{\overline{\varepsilon}_{KK}(\omega)\right\}&=1+\frac{2}{\pi}\fint_{0}^{\infty}
\frac{\omega^{\prime}\mathfrak{Im}\left\{\overline{\varepsilon}(\omega^{\prime})\right\}}{\omega^{\prime2}-\omega^{2}}d\omega^{\prime}, \\
\label{eq:ImKKR}
\mathfrak{Im}\left\{\overline{\varepsilon}_{KK}(\omega)\right\}&=-\frac{2}{\pi}\fint_{0}^{\infty}
\frac{\omega^{\prime}[\mathfrak{Re}\left\{\overline{\varepsilon}(\omega^{\prime})\right\}-1]}{\omega^{\prime2}-\omega^{2}}d\omega^{\prime},
\end{align}
\end{subequations}
where the symbol $\fint$ denotes the Cauchy principal value of the integral. Note that for normal
incidence $\overline{\varepsilon}$ does not depend on the polarization angle \cite{rpa07ol}, and
$\overline{\varepsilon}_{xx}=\overline{\varepsilon}_{yy}$ and
$\overline{\varepsilon}_{xy}=\overline{\varepsilon}_{yx}=0$.

Figure~\ref{fig:HomoSuscepbilityFF} reveals several important properties of the wavelength
dispersion of the effective permittivity of the homogenized metasurface. First, it is markedly
different from the graphene permittivity and shows a series of Lorentz-type resonances, which
correspond to localized surface plasmons of the graphene crosses. Moreover, whereas
$\mathfrak{Re}(\varepsilon_{g})<0$ in the entire wavelength domain we considered, in some
wavelength domains $\mathfrak{Re}(\overline{\varepsilon})>0$. This suggests that at those
wavelengths the metasurface responds as a dielectric one. Equally important, it can be seen in
Fig.~\ref{fig:HomoSuscepbilityFF} that the permittivity $\overline{\varepsilon}_{KK}$ obtained from
the effective permittivity of the metasurface, $\overline{\varepsilon}$, using the KK relations is
almost identical with the latter, $\overline{\varepsilon}_{KK}\approx\overline{\varepsilon}$, which
validates our method. More specifically, the effective permittivity calculated with our
homogenization method obeys the causality principle, proving that it is a physically meaningful
quantity.
\begin{figure}[t!]\centering
\includegraphics[width=8.5 cm]{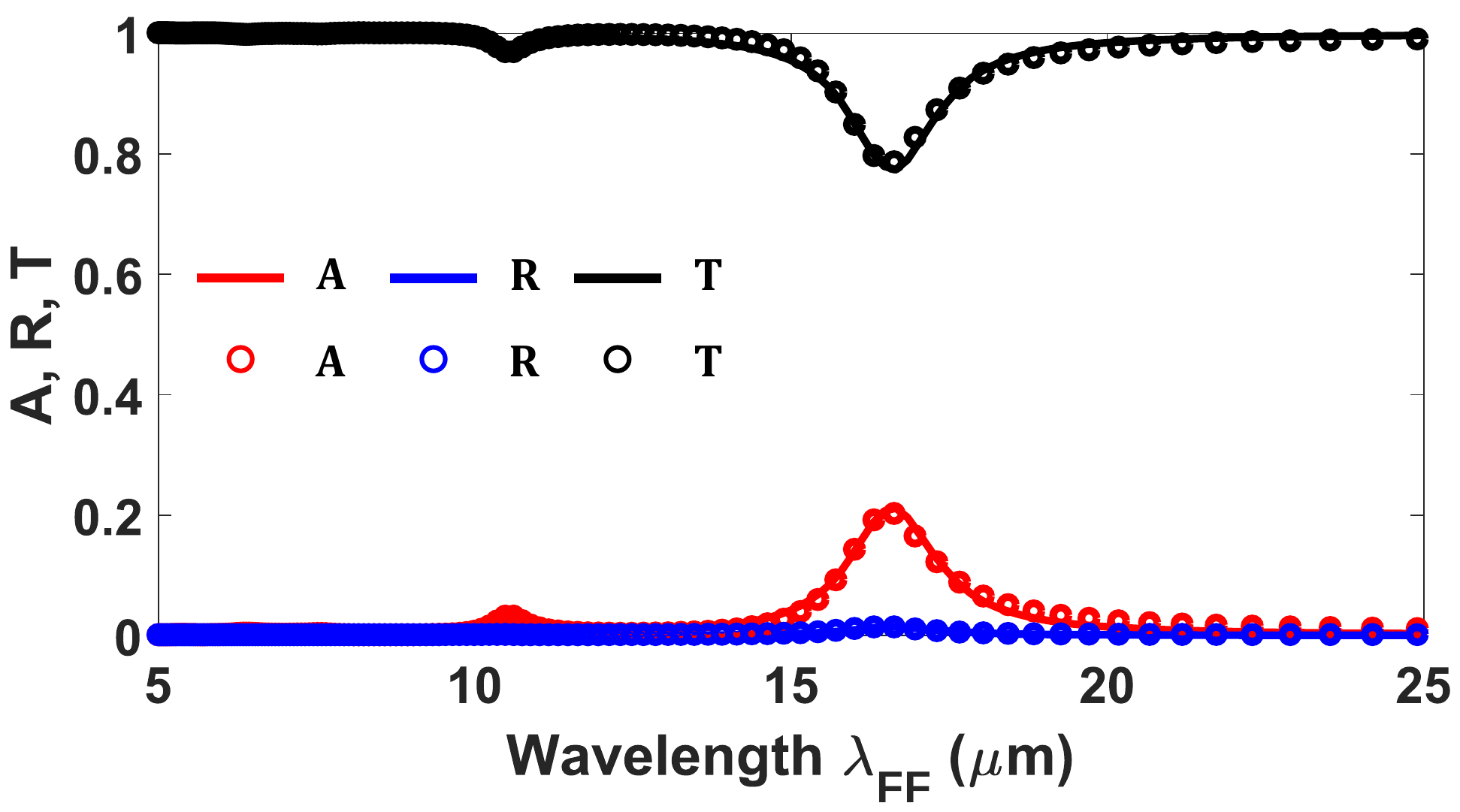}
\caption{Comparison between the absorption, $A$, reflectance, $R$, and transmittance, $T$, of the
original cruciform graphene metasurface (solid curves) and the homogenized one (circles).}
\label{fig:HomoResponseFF}
\end{figure}

A key advantage of the homogenization theory is that a metasurface patterned in a convolute manner
can be replaced with a simple homogenous layer of a material with specific optical constants in
such a way that physical quantities such as absorption, $A$, reflectance, $R$, and transmittance,
$T$, of the two optical systems are identical. This provides us an effective tool to validate our
homogenization method, namely we quantify the difference between the absorption, reflectance, and
transmittance of the original metasurface and the homogenized one. To this end, we have computed
these quantities for both optical systems using the FDTD method. The results of these calculations
are summarized in Fig.~\ref{fig:HomoResponseFF}. This comparison clearly shows that the linear
response of the uniform layer of material with the retrieved effective permittivity is practically
the same as that of the graphene cruciform metasurface. This excellent agreement further proves the
reliability of our linear homogenization method. Moreover, the data presented in
Fig.~\ref{fig:HomoResponseFF} suggest that the peaks of the absorption spectra shown in this figure
coincide with those of the imaginary part of the retrieved permittivity plotted in
Fig.~\ref{fig:HomoSuscepbilityFF}. This finding is explained by the fact that the peaks in the
spectrum of the effective permittivity of the homogenized metasurface correspond to the excitation
of localized surface plasmons on the graphene crosses, a phenomenon accompanied by large
enhancement of the optical near-field and consequently an increase of the optical absorption.

Encouraged by the accuracy with which our homogenization method describes the linear optical
response of graphene metasurfaces, we proceeded to extend it to the much more complex case of
nonlinear optical interactions. Due to the centrosymmetric nature of the graphene lattice, the
lowest-order non-vanishing nonlinear optical interactions in graphene are of the third-order. In
particular, in the case of the third-harmonic generation (THG), the nonlinear polarization is given
by:
\begin{equation}\label{eq:ConstTH}
\mathbf{P}^{nl}(\mathbf{r},\Omega)=
\varepsilon_{0}\bm{\chi}^{(3)}(\mathbf{r},\Omega,\omega)\vdots\mathbf{E}(\mathbf{r},\omega)\mathbf{E}(\mathbf{r},\omega)\mathbf{E}(\mathbf{r},\omega),
\end{equation}
where $\bm{\chi}^{(3)}$ is the third-order susceptibility and $\Omega=3\omega$ is the frequency of
the third harmonic (TH). Componentwise, this nonlinear polarization can be written as:
\begin{equation}\label{eq:CompP}
P^{nl}_{i}=\varepsilon_{0}\sum_{jkl}\chi_{ijkl}^{(3)}{E_j}{E_k}{E_l}\equiv\varepsilon_{0}\sum_{jkl}q_{ijkl},
\end{equation}
where the nonlinear auxiliary quantities $q_{ijkl}=\chi_{ijkl}^{(3)}{E_j}{E_k}{E_l}$ have been
defined. Their averaged values are:
\begin{equation}\label{eq:AverQ}
\overline{q}_{ijkl}(\Omega)=\frac{1}{V}\int_V\chi_{ijkl}^{(3)}(\mathbf{r},\Omega,\omega)
E_{j}(\mathbf{r},\omega)E_{k}(\mathbf{r},\omega)E_{l}(\mathbf{r},\omega)d\mathbf{r}.
\end{equation}
\begin{figure}[t!]\centering
\includegraphics[width=8.5 cm]{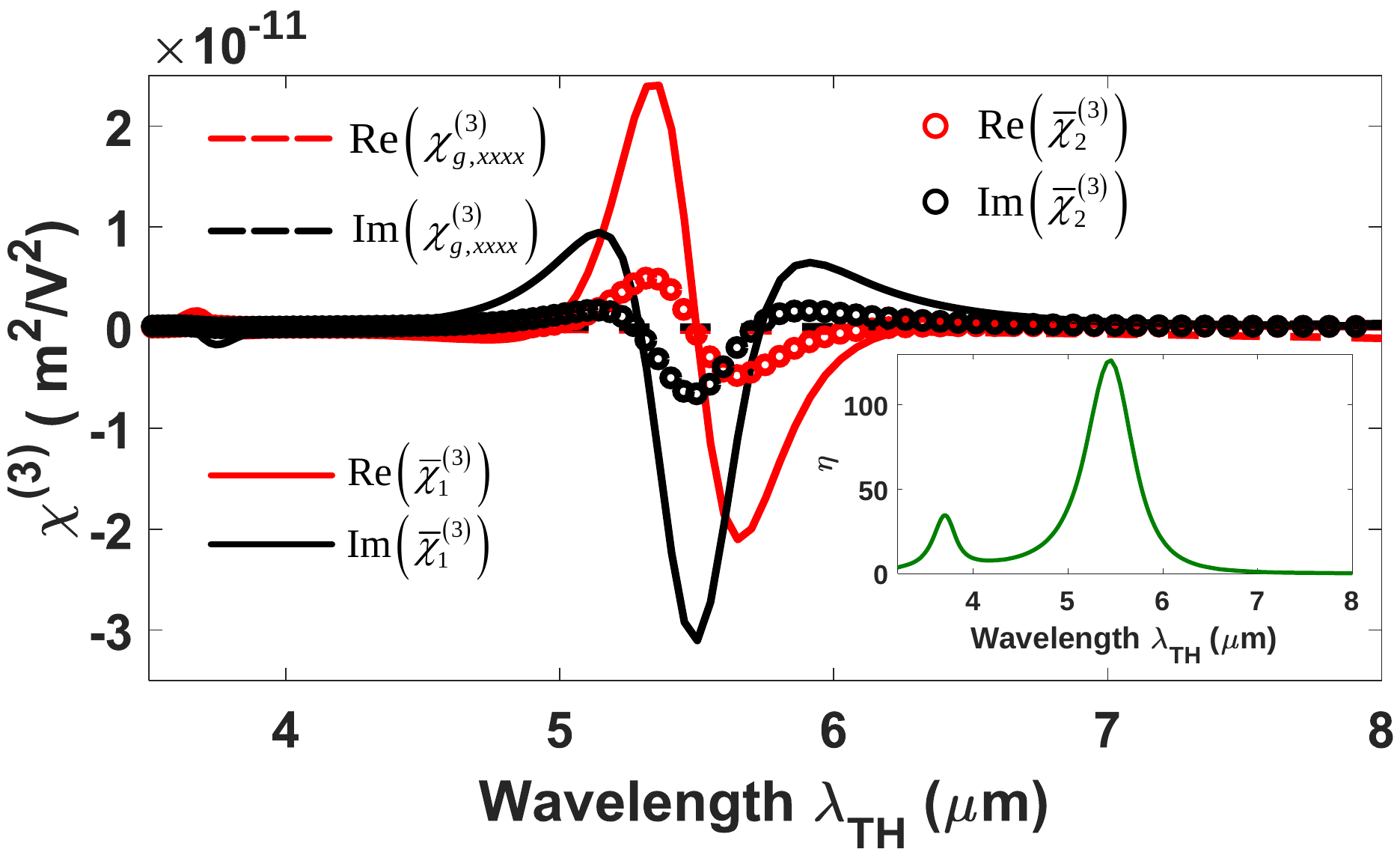}
\caption{Wavelength dependence of the dominant component of the third-order susceptibility of
graphene, $\chi_{g,xxxx}^{(3)}$, and of the two independent components of the effective third-order
susceptibility of the graphene metasurface, $\overline{\chi}_{1}^{(3)}$ and
$\overline{\chi}_{2}^{(3)}$. In inset, the spectrum of the enhancement factor
$\eta=\vert\overline{\chi}_{1}^{(3)}\vert/\vert\chi_{g}^{(3)}\vert$.}
\label{fig:HomoSuscepbilityTH}
\end{figure}

Similarly to \eqref{eq:CompP}, the nonlinear polarization in the homogenized layer can be expressed
in terms of the averaged fields at the fundamental frequency (FF) and an effective third-order
susceptibility, $\overline{\bm{\chi}}^{(3)}$, as:
\begin{equation}\label{eq:CompPav}
\overline{P}^{nl}_{i}(\Omega)=\varepsilon_{0}\sum_{jkl}\overline{\chi}_{ijkl}^{(3)}(\Omega,\omega)\overline{E}_j(\omega)\overline{E}_k(\omega)\overline{E}_l(\omega).
\end{equation}

Finally, we require that the averaged nonlinear polarizations described by \eqref{eq:CompP} and
\eqref{eq:CompPav} are \textit{termwise} identical. This ensures that on average the nonlinear
polarizations in the graphene metasurface and the homogenized layer of nonlinear material are
equal. Under these circumstances, the effective third-order susceptibility,
$\overline{\bm{\chi}}^{(3)}$, is given by the following formula:
\begin{equation}\label{eq:AverChi}
\overline{\chi}_{ijkl}^{(3)}(\Omega,\omega)=\frac{\overline{q}_{ijkl}(\Omega)}{\overline{E}_j(\omega)\overline{E}_k(\omega)\overline{E}_l(\omega)}.
\end{equation}

We have used this formalism, namely, Eqs.~(\ref{eq:AverQ}) and (\ref{eq:AverChi}) in conjunction
with \eqref{eq:AverE}, to compute the effective third-order susceptibility
$\overline{\bm{\chi}}^{(3)}$ of the graphene cruciform metasurface, and summarize the relevant
results in Fig.~\ref{fig:HomoSuscepbilityTH}. The third-order susceptibility of a homogeneous
graphene sheet is $\bm{\chi}^{(3)}_{g}(3\omega;\omega)=[i/(3\omega \varepsilon_0
h_{g})]\bm{\sigma}_s^{(3)}(3\omega;\omega)$, where the third-order conductivity is given by
$\sigma_{s,ijkl}^{(3)}(3\omega;\omega)=\sigma_{s}^{(3)} (\delta_{ij}\delta_{kl} +
\delta_{ik}\delta_{jl} + \delta_{il}\delta_{jk})/3$ \cite{csa16acsnano}, where $\delta_{ij}$ is the
Kronecker delta and $\sigma_s^{(3)} = \frac{i\sigma_0{\left(\hbar v_F e\right)}^2}{48\pi
{\left(\hbar\omega\right)}^4}T\left(\frac{\hbar\omega}{2\vert\mu_c\vert} \right)$. Here, $v_F
\approx c/300$ is the Fermi velocity, $\sigma_0 = e^2/(4\hbar)$ is the universal dynamic
conductivity of graphene, and $T(x) = 17G(x) - 64G(2x) + 45G(3x)$, where
$G(x)=\ln\vert(1+x)/(1-x)\vert+i\pi H(|x|-1)$ and $H(x)$ is the Heaviside step function.
\begin{figure}[t!]
\centering
\includegraphics[width=8.5 cm]{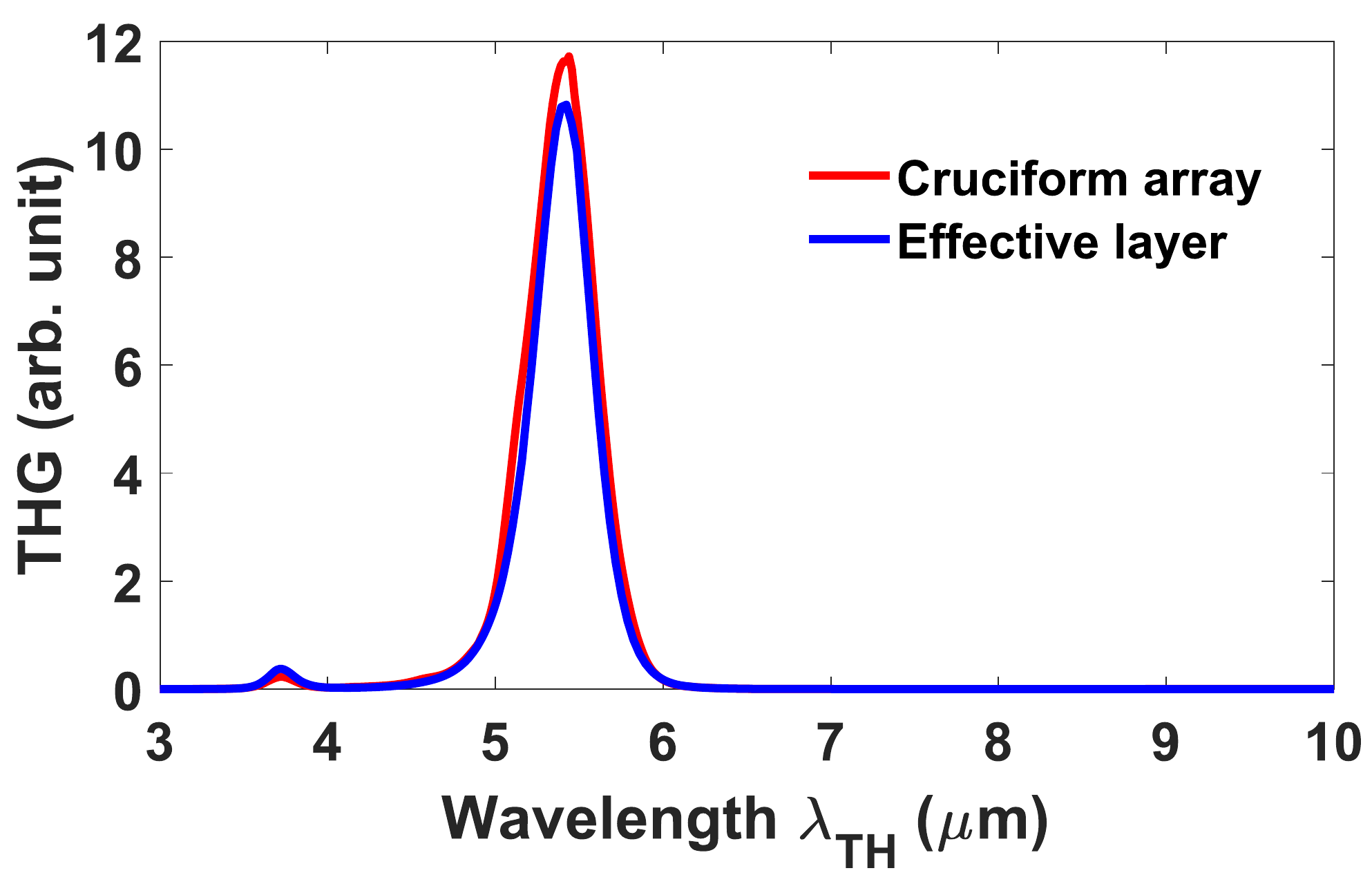}
\caption{Third-harmonic generation spectra calculated for a graphene cruciform metasurface and the
corresponding homogenized layer of nonlinear optical material.} \label{fig:HomoResponseTH}
\end{figure}

The symmetry group of graphene lattice is $D_{6h}$, so that $\bm{\chi}_{g}^{(3)}$ has only two
independent non-zero components: $\chi_{g,xxxx}^{(3)}=\chi_{g,yyyy}^{(3)}$ and
$\chi_{g,xyxy}^{(3)}=\chi_{g,xxyy}^{(3)}=\chi_{g,xyyx}^{(3)}=\chi_{g,yxxy}^{(3)}=\chi_{g,yxyx}^{(3)}=\chi_{g,yyxx}^{(3)}$.
The dominant component is $\chi_{g,xxxx}^{(3)}\equiv\chi_{g}^{(3)}$, and is depicted by dashed
curves in Fig.~\ref{fig:HomoSuscepbilityTH}. Importantly, it can be readily seen from
\eqref{eq:AverChi} that $\overline{\chi}_{ijkl}^{(3)}$ and $\chi_{g,ijkl}^{(3)}$ have the same set
of non-zero components, so that the effective third-order susceptibility has only two independent
non-zero components, too: $\overline{\chi}_{xxxx}^{(3)}\equiv\overline{\chi}_{1}^{(3)}$ and
$\overline{\chi}_{xyxy}^{(3)}\equiv\overline{\chi}_{2}^{(3)}$. In order to quantify the enhancement
of the nonlinear optical response of the graphene metasurface, we also computed the ratio
$\eta=\vert\overline{\chi}_{1}^{(3)}\vert/\vert\chi_{g}^{(3)}\vert$.

The results presented in Fig.~\ref{fig:HomoSuscepbilityTH} reveal several important conclusions.
First, the dominant component of $\overline{\bm{\chi}}^{(3)}$ is $\overline{\chi}_{xxxx}^{(3)}$,
but unlike the monotonous frequency dependence of graphene third-order susceptibility, the
frequency dependence of $\overline{\chi}_{xxxx}^{(3)}$ suggests a resonant nonlinear optical
response. As in the linear case, the resonances of the optical nonlinearity of the homogenized
metasurface coincide with the plasmon-induced peaks shown in Fig.~\ref{fig:HomoSuscepbilityFF}.
Second, due to these plasmon-induced resonances, the effective third-order susceptibility of the
graphene metasurface is strongly enhanced as compared to that of a graphene sheet; \textit{e.g.},
at the wavelength $\lambda_{TH}=\SI{5.4}{\micro\meter}$ of the main resonance the enhancement
factor $\eta$ is about two orders of magnitude.

In order to validate the nonlinear part of the proposed homogenization method, the THG of both the
graphene cruciform metasurface and the corresponding homogeneous layer of material have been
calculated using an in-house developed generalized-source-FDTD code \cite{yp17tap}. The results of
these computations are presented and compared in Fig.~\ref{fig:HomoResponseTH}, where the
corresponding THG spectra are plotted. It can be seen from this figure that there is a very good
agreement between the two spectra, a maximum difference of about \SI{5}{\percent} being observed at
the wavelength of the main resonance. Moreover, the very good agreement regarding both linear and
nonlinear response of the graphene metasurface and its homogenized counterpart, illustrated in
Fig.~\ref{fig:HomoResponseFF} and Fig.~\ref{fig:HomoResponseTH}, respectively, indicate that a
patterned graphene metasurface can be accurately replaced with a simple homogenous layer of
material with an effective permittivity and nonlinear susceptibility retrieved with our
homogenization method.

In summary, we have introduced a novel and efficient homogenization method for the analysis of the
linear and nonlinear optical response of graphene metasurfaces. Our study shows that the
third-order nonlinearity of such metasurfaces is enhanced by more than two order of magnitude at
the resonance frequencies of surface plasmons of the graphene components of the metasurface. Due to
its versatility, our method can be extended to metasurfaces containing two-dimensional materials
other than graphene, to three-dimensional metamaterials, and to a large class of nonlinear optical
interactions. Due to all these powerful features, our method could have great potential to
facilitate the design of active photonic devices with advanced functionalities.

\textit{Funding} -- H2020 European Research Council (ERC) (ERC-2014-CoG-648328).

\end{document}